\documentclass[twocolumn,pra,aps,superscriptaddress,amsmath,amssymb]{revtex4-1}%,floatfix
\usepackage{color}
\usepackage{xcolor}
\usepackage{mathrsfs}
\usepackage{amsmath}
\usepackage{graphicx}% Include figure files
\usepackage{dcolumn}% Align table columns on decimal point
\usepackage{bm}% bold math
\usepackage{times}
\usepackage{amssymb}
\usepackage{float}
\usepackage{array}
\usepackage{appendix}
\usepackage{color}
\usepackage[colorlinks,citecolor=blue]{hyperref}

\begin{document}
	
\def\sgn{\mathop{\rm sgn}}
		
\title{Faithful Simulation and Detection of Quantum Spin Hall Effect on Superconducting Circuits}

\author{Jia Liu}%\email{liuj.phys@foxmail.com}
\affiliation{Guangdong Provincial Key Laboratory of Quantum Engineering and Quantum Materials,
and School of Physics and Telecommunication Engineering,
South China Normal University, Guangzhou 510006, China}
\affiliation{Guangdong-Hong Kong Joint Laboratory of Quantum Matter, and Frontier Research Institute for Physics, South China Normal University, Guangzhou 510006, China}

\author{Jun-Yi  Cao}%\email{793092733@qq.com}
\affiliation{Guangdong Provincial Key Laboratory of Quantum Engineering and Quantum Materials,
and School of Physics and Telecommunication Engineering,
South China Normal University, Guangzhou 510006, China}

\author{Gang Chen}  \email{chengang971@163.com}
\affiliation{State Key Laboratory of Quantum Optics and Quantum Optics Devices,
	Institute of Laser Spectroscopy, Shanxi University, Taiyuan, Shanxi 030006, China}
\affiliation{Collaborative Innovation Center of Extreme Optics, Shanxi University,
Taiyuan 030006, China}
\affiliation{Collaborative Innovation Center of Light Manipulations and Applications, Shandong Normal University, Jinan 250358, China}
	
\author{Zheng-Yuan Xue}  \email{zyxue83@163.com}
\affiliation{Guangdong Provincial Key Laboratory of Quantum Engineering and Quantum Materials,
and School of Physics and Telecommunication Engineering,
South China Normal University, Guangzhou 510006, China}
\affiliation{Guangdong-Hong Kong Joint Laboratory of Quantum Matter, and Frontier Research Institute for Physics, South China Normal University, Guangzhou 510006, China}

\date{\today}

\begin{abstract}
Topological states of quantum matter have inspired both fascinating physics findings and exciting opportunities for applications. Due to the over-complicated structure of, as well as interactions between, real materials, a faithful quantum simulation of topological matter is very important in deepening our understanding of these states. This requirement puts the quantum superconducting circuits system as a good option for mimicking topological materials, owing to their flexible tunability and fine controllability. As a typical example herein, we realize a $\mathbb{Z}_2$-type topological insulator featuring the quantum spin Hall effect state, using a coupled system of transmission-line resonators and transmons. The single-excitation eigenstates of each unit cell are used as a pseudo-spin 1/2 system. The boundary of the topological phase transition is fixed in the phase diagram. Topological edge states are shown, which can be experimentally verified by detecting the population at the boundary of the plane. Compared to the previous simulations, this compositional system is fairly controllable, stable and less limited. Therefore, our scheme provides a reliable platform for faithful quantum simulations of topological matter.
\end{abstract}

\maketitle

\section{Introduction}

The discovery of topological matters is a triumph of solid state physics, and since the fractional quantum hall effect was
disclosed in 1980s, many theoretical and experimental efforts have been made
regarding topological matters. In particularly, %the very beginning of 21 century,
a new type insulator has been predicted, which has edge current along the surface, but be insulated in the bulk, i.e., the so called quantum spin Hall effect (QSHE) \cite{Kane_Z2,Kane_TI,Zhang_QSH,Zhang_QSH2,Qi_QSHTI}. Different from the quantum Hall effect, the QSHE has time reversal symmetry and no external magnetic field is needed for its realization. In 2007, the QSHE was discovered in 2D topological insulator HgTe/CdTe system \cite{QSH_exp}. %which was predicted in Ref.~\cite{Zhang_QSH}.
From then on, many experiments were carried out for the realization of virous types of topological matters \cite{tm1,tm2,tm3,tm4}. Although many topological materials had been predicted theoretically, the realization %of these proposals
is very little because such natural materials are still very lacking. Meanwhile, the complicated structure and fixed interactions in real materials prevent us from analytical investigating the physical origin. Therefore, simulating topological properties in well controllable physical systems still has great significance for exploring more deep-seated topological phenomena and enhancing our understanding of the role of topology in quantum materials.

In the past two decades, many quantum simulation schemes for topological matters have been suggested in the ultracold atoms \cite{TV3, tm5, coldatoms_1, coldatoms_2, coldatoms_3, Hamiltonian1, coldatoms_4, coldatoms_5,coldatoms_6} and optical system \cite{coldatoms_7, tm6,Xue_npj,phonon_1,phonon_2}. But it is not easy to simulate topological matters in both atomic and optical systems. In atomic system, it is difficult to implement individual control and the same reason makes the detection of the topological phenomena to be hard neither. For optical system, there is always limitations of the interactions, e.g., the hopping phase can not be controlled freely in experiment \cite{Hamiltonian1}. In addition, in optical systems, most proposals concern with spinless bosonic system  and  not suitable for simulating spinful systems.

Recently, superconducting circuits \cite{cqed1,cqed2,cqed3,Nori-rew-Simu2-JC}, a scalable  quantum computation platform,
has been applied to simulate quantum many-body systems \cite{s1,s2,s3,s4,s5,s6,s7,s8,s9,TS1D_prl}. In the spin concern system \cite{cqed4,Xue_prappl,TS1D_2}, the hopping between each lattice can be adjusted separately and there is no limitation on nearly all parameters that are used to fix the physical properties of the system, such as the hopping strength, on-site potential, hopping phase, etc. Here, we propose a coupled transmission-line resonators (TLRs) and transmons system to simulate a 2D topological insulator. %We vary the coupling coefficients and hopping phases of the system, fix the parameters where the system is the topological insulator. Phase diagram is fixed to show where the QSHE can be realized. We also give out the edge states' distribution %along the 2D plane
%for special cases, which can be detected by resonance absorption experimentally.

\section{Simulation of the QSHE}
\begin{figure}[tb]
\center
\includegraphics[width=0.9\linewidth]{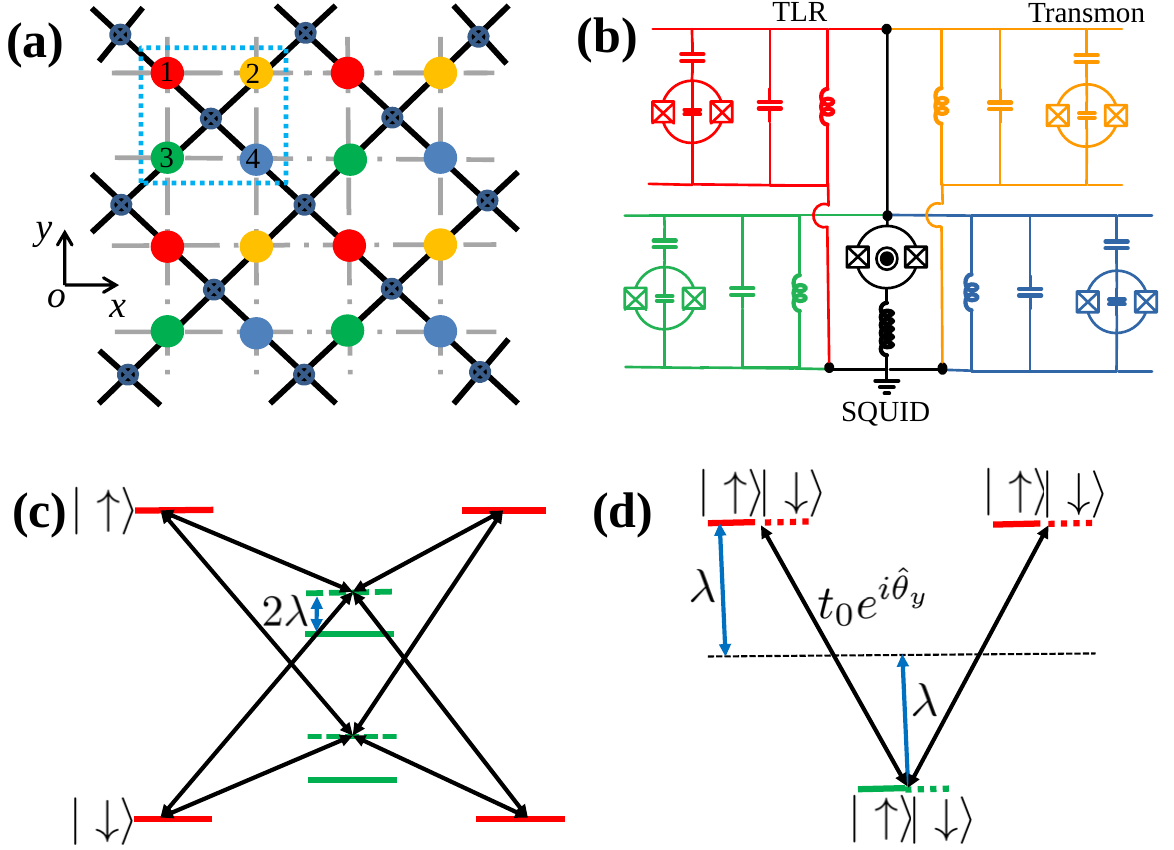}
\caption{ (a) The proposed 2D JC lattice with each unit contains four sites, illustrated by different colors, with one SQUID connected them, as detailed in (b). (b) Each site consists of  one transmon coupled to a TLR, the hopping between neighboring sites is realized by the SQUID. Different Peierls phases of the hoppings in $x$ and $y$-directions can be realized by setting appropriate parameters in the corresponding lattice and the SQUID. (c) Taking out one column in (a) for explanation. For the hopping between the lattices in $y$-direction 1-3-1, set the detuning in the No. 3 of each two lattices in $y$-direction, after unitary transformation we can get the staggered on-site potential as shown in (d). (d) Interacted lattices with staggered on-site potential along $y$-direction.}
  \label{fig1}
\end{figure}

As shown in Fig.~\ref{fig1}(a), we consider a 2D lattice with the following model Hamiltonian \cite{Hamiltonian1}
\begin{eqnarray}\label{hami}
\mathcal{H}&=&-t_0\sum_{m,n}{\bf{c}}_{m+1,n}^\dag e^{i\hat{\theta}_x}{\bf{c}}_{m,n}+{\bf{c}}_{m,n+1}^\dag e^{i\hat{\theta}_y}{\bf{c}}_{m,n}+\text{H.c.}\notag\\
&&+\sum_{m,n}\lambda_{m,n}{\bf{c}}_{m,n}^\dag {\bf{c}}_{m,n},
\end{eqnarray}
where $t_0$ is the nearest-neighbor hopping strength; ${\bf{c}}_{m,n}=(c_{m,n,\uparrow},c_{m,n,\downarrow})^T$ is a 2-component fermi operator defined on a lattice site ($x=ma, y=nb$) with $a$ and $b$ being the lattice spacings and $m$ and $n$ being integers; $\hat{\theta}_x=2\pi\alpha y\sigma_z$ and $\hat{\theta}_y=2\pi\beta \sigma_x$ with $(\sigma_x,\sigma_z)$ being the Pauli matrices and $(\alpha,\beta)$ being parameters governed by the magnetic flux and spin mixing; $\lambda_{m,n}$ is the on-site potential which is here set to be staggered in $y$-direction, i.e., $\lambda_{m,n}=(-1)^n\lambda$. Define the time reversal operator as $\mathcal{T}$$=i\sigma_y K$, where $\sigma_y$ is also the Pauli matrix and $K$ denotes the complex conjugation. We can prove that $\mathcal{T}$ commutes with the Hamiltonian in Eq.~(\ref{hami}),so the Hamiltonian in Eq.~(\ref{hami}) has the time reversal symmetry. This system belongs to topological class with topological index $\mathbb{Z}_2$ \cite{z2_class} and can be used to realize the QSHE.

Here we first take one rectangle block including four sites to introduce how to achieve the Hamiltonian in Eq.~(\ref{hami}). As shown in Fig.~\ref{fig1}(a), each circle presents a Jaynes-Cummings (JC) model made of one TLR and transmon, and the lattices are connected by a superconducting quantum interference device (SQUID) and a inductor $L$ \cite{Nori-rew-Simu2-JC}. The hoppings between nearest-neighbor lattices with Peierls phases $\hat{\theta}_{x,y}$ can be adjusted by  tuning the magnetic field through the connected SQUID. The hoppings in $x$ and $y$-directions are independent from each other and can be realized in the similar way. The Hamiltonian of that JC lattice is
\begin{equation}
	\mathcal{H}_{\text{JC}}=\sum_{\textbf{r}}h_{\textbf{r}}+\sum_{\langle\textbf{r}\textbf{r}'\rangle
}J_{\textbf{r}\textbf{r}'}(t)\left(\hat{a}_{\textbf{r}}^\dagger \hat{a}_{\textbf{r}'}+\text{H.c.}\right), \label{H_JC}
\end{equation}
where $\textbf{r}$ is the label of the unit cell at $(x,y)$; $h_{\textbf{r}}= \hbar \omega_{\textbf{r}} (\sigma_{\textbf{r}}^+\sigma_{\textbf{r}}^- + \hat{a}_{\textbf{r}}^\dagger \hat{a}_{\textbf{r}}) + g_{\textbf{r}}(\sigma_{\textbf{r}}^+ \hat{a}_{\textbf{r}} + \sigma_{\textbf{r}}^- \hat{a}_{\textbf{r}}^\dagger)$ is the JC Hamiltonian in unit cell at $\textbf{r}$; $\sigma_{\textbf{r}}^{+}=|e\rangle_{\textbf{r}}\langle g|$ and $\sigma_{\textbf{r}}^{-}=|g \rangle_{\textbf{r}}\langle e|$ are the raising and lowering operators of the transmon qubit at $\textbf{r}$; $\hat{a}_{\textbf{r}}$ and $\hat{a}_{\textbf{r}}^\dagger$ are the annihilation and creation operators of the photon in the TLR at $\textbf{r}$ with the frequency of $\omega_{\textbf{r}}$ and the condition $g_{\textbf{r}}\ll\omega_{\textbf{r}}$ has to be met for justifying the JC coupling; $J_{\textbf{r}\textbf{r}'}(t)$ is the inter-cell hopping strength between the unit cell in $\textbf{r}$ and its neighbor cells. In the following of the paper, for each ~$\textbf{r}'=(ma,nb)$ we set $\textbf{r}=(ma,(n+1)b)$ or $\textbf{r}=((m+1)a,nb)$. The lowest three eigenstates of the JC Hamiltonian $h_{\textbf{r}}$ are denoted as $|0g\rangle_{\textbf{r}}$, $|$$\uparrow\rangle_{\textbf{r}} =\left(|0e\rangle_{\textbf{r}} + |1g\rangle_{\textbf{r}} \right)/\sqrt{2}$ and $|$$\downarrow \rangle_{\textbf{r}} =\left(|0e\rangle_{\textbf{r}} - |1g\rangle_{\textbf{r}} \right)/\sqrt{2}$, where $|ng\rangle_{\textbf{r}}$ and $|ne\rangle_{\textbf{r}} \; (n=0, 1, 2, \ldots)$ are the states containing $n$ photons while the transmon is at the ground and excited states. The corresponding eigen-energies are $E_{\textbf{r},0g}=0$ and $E_{\textbf{r},\uparrow/\downarrow}= \omega_{\textbf{r}} \pm g_{\textbf{r}}$.  We choose the two single-excitation eigenstates $|\uparrow\rangle_{\textbf{r}}$ and $|\downarrow\rangle_{\textbf{r}}$ to simulate the effective spin-up and spin-down states in the lattice at $\textbf{r}$.  We can control each hopping separately by adjusting the pulse shape of $J_{\textbf{r}\textbf{r}'}(t)$.  Based on the current experiments \cite{cqed1}, setting $t_0/2\pi=3$ MHz, $\omega_{\text{1}}/t_0=2700$, $\omega_{\text{2}}/t_0=3000$, $\omega_{\text{3}}/t_0=2650$, and $\omega_{\text{4}}/t_0=2900$. And $g_{\text{1}}/t_0=250$, $g_{\text{2}}/t_0=150$, $g_{\text{3}}/t_0=150$, $g_{\text{4}}/t_0=200$. With those, the energy interval $|E_{{\textbf{r}},\eta}-E_{{\textbf{r}'},\eta'}|$ of the 16 hopping between each two of them is much larger than (or equal to) $20$ times of the effective hopping strength $t_0$, such distance is enough for selective frequency addressing. And for $16$ tunes the hopping strength
	\begin{equation}\label{coupling2}
	J_{\textbf{r}\textbf{r}'}(t)=\sum_{\eta, \eta'} 4t_{0,\textbf{r}\textbf{r}'\eta\eta'} \cos\left(\omega_{\textbf{r}\textbf{r}'\eta\eta'} t+  s_{\textbf{r}\textbf{r}'\eta\eta'} \varphi_{\textbf{r}\textbf{r}'\eta\eta'}\right),
	\end{equation}
where $s_{\textbf{r}\textbf{r}'\eta\eta'}= \text{sgn}(E_{\textbf{r},\eta}-E_{\textbf{r}',\eta'})$ is the sign of the hopping phase, and ~$\omega_{\textbf{r}\textbf{r}'\eta\eta'}=|E_{\textbf{r},\eta}-E_{\textbf{r}',\eta'}|$ is the energy difference between the nearest lattice. $4t_{0,\textbf{r}\textbf{r}'\eta\eta'}$ and $s_{\textbf{r}\textbf{r}'\eta\eta'} \varphi_{\textbf{r}\textbf{r}'\eta\eta'}$ are the amplitude and phase corresponding to the hopping $|\eta\rangle_{\textbf{r}} \rightarrow |\eta'\rangle_{\textbf{r}'}$, respectively. In experiments, this time-dependent coupling strength $J_{\textbf{r}\textbf{r}'}(t)$ can be realized by adding external magnetic fluxes with dc and ac components through the SQUIDs \cite{TS1D_2}. Both the hopping strengths and phases can be controlled by inducing controllable spin transition process in a certain rotating frame via Eq.~(\ref{coupling2}).

%\subsection{Engineering the hopping}

We proceed to show how the selective control of individual hopping in the JC lattice can be achieved by adjusting $J_{\textbf{r}\textbf{r}'}(t)$ in  Eq.~(\ref{coupling2}) via the ac flux. First mapping the Hamiltonian in Eq.~(\ref{H_JC}) into the single excitation subspace by using $|\eta \rangle_{\textbf{r}}$ to denote one excitation state with 'spin' ~$\eta=\uparrow,\downarrow$ and get the Hamiltonian in the dressed states,
\begin{equation}
\begin{aligned}
\mathcal{H}_{\text{JC}}^{S}
=\sum_{{\textbf{r}},\eta} E_{{\textbf{r}},\eta}|\eta\rangle_{{\textbf{r}}} \langle\eta|+\frac{1}{2}\sum_{{\textbf{r}}\textbf{r}',\eta\eta'}J_{\textbf{r}\textbf{r}'}(t) |\eta \rangle_{{\textbf{r}},{\textbf{r}'}}\langle\eta'|+\text{H.c.}.
\end{aligned}
\end{equation}
Then for each $J_{\textbf{r}\textbf{r}'}(t)$, we add four tunes, each in resonant to one of the $16$ inter-cell hoppings \cite{TS1D_2}, and contains its independent tunable amplitude, frequency and phase as shown in Eq.~(\ref{coupling2}). The form of $J_{\textbf{r}\textbf{r}'}(t)$ will be determined by $t_{0,\textbf{r}\textbf{r}'\eta\eta'}$ and $\varphi_{{\textbf{r}}\textbf{r}'\eta\eta'}$ depending on the topological phase in simulated scheme. The target Hamiltonian we need to simulate topological insulator can be got in the rotating frame transformed by
%\begin{equation}
$U=\exp\{-i\left[\sum_{{\textbf{r}}} h_{\textbf{r}}-(-1)^{n_{\textbf{r}}}\lambda(|\uparrow\rangle_{\textbf{r}}\langle \uparrow|+ |\downarrow\rangle_{\textbf{r}}\langle \downarrow | )\right] t\}$,
%\end{equation}
where $n_{\textbf{r}}$ is the same number as $n$ in the ${\textbf{r}}=(x,y)=(ma,nb)$. After the unitary transformation $\mathcal{H}'_{\text{JC}}=U^{\dag}\mathcal{H}_{\text{JC}}U+i\dot{U}^{\dag}U$, and if the conditions $\{ t_{0,{\textbf{r}}\textbf{r}'\eta\eta'}\}_{\eta,\eta'=\uparrow,\downarrow}\ll  \{ \omega_{{\textbf{r}}\textbf{r}'\eta\eta'},\; \omega_{{\textbf{r}}\textbf{r}'\eta\eta'}- \omega_{{\textbf{r}}\textbf{r}'\bar{\eta}\bar{\eta}'} \}_{\eta,\eta',\bar{\eta},\bar{\eta}' =\uparrow,\downarrow; \; \omega_{{\textbf{r}}\textbf{r}'\eta\eta'}\ne \omega_{{\textbf{r}}\textbf{r}'\bar{\eta}\bar{\eta}'} }$ are satisfied, all the other terms are fast rotating term that can be dropped with the rotating wave approximation. As a result, we derive the 2D tight-binding model with tunable hopping coefficients as
\begin{eqnarray} \label{eq.simu}
\mathcal{H}_{\text{TB}}
&=&\sum_{\langle{\textbf{r}}{\textbf{r}'}\rangle}\sum_{\eta, \eta'}  t_{0,{\textbf{r}}\textbf{r}'\eta\eta'}e^{i\varphi_{{\textbf{r}}\textbf{r}'\eta\eta'}}
\hat{c}^\dagger_{{\textbf{r}},\eta} \hat{c}_{{\textbf{r}}',\eta'}+\text{H.c.} \notag\\
&&+\sum_{{\textbf{r}}}(-1)^n\lambda\sigma^{0}_{\textbf{r}},
\end{eqnarray}
where $\,t_{0,{\textbf{r}}\textbf{r}'\eta\eta'}$ is just the effective coupling strength and $\hat{c}^\dagger_{{\textbf{r}},\eta}=|\eta\rangle_{\textbf{r}} \, \langle G | $ is the creation operator of electron with `spin' $\eta$ in the lattice at ${\textbf{r}}$. Though the superconducting qubit has bosonic nature,the interaction of quasi-fermions between each lattice is totally the same as the real electrons, in such subspace the time reversal symmetry can be kept as well as the Hamiltonian in Eq. (1). Comparing the Hamiltonian in Eqs.~(\ref{hami}) and ~(\ref{eq.simu}), we should choose appropriate $\,t_{0,{\textbf{r}}\textbf{r}'\eta\eta'}$ and $\varphi_{{\textbf{r}}\textbf{r}'\eta\eta'}$ to get the target model in Eq.~(\ref{hami}).
%expressions in sentence, after discussion will left one version
%In $x$-direction, choose $t_{0,((m+1),n),(m,n),\uparrow\uparrow}=t_{0,((m+1),n),(m,n),\downarrow\downarrow}=t_0$ and $\varphi_{((m+1),n),(m,n),\uparrow\uparrow}=2n\pi\alpha$, $\varphi_{((m+1),n),(m,n),\downarrow\downarrow}=-2n\pi\alpha$. In $y$-direction, set $t_{0,(m,(n+1)),(m,n),\uparrow\uparrow}=t_{0,(m,(n+1)),(m,n),\uparrow\downarrow}=t_{0,(m,(n+1)),(m,n),\downarrow\uparrow}=t_{0,(m,(n+1)),(m,n),\downarrow\downarrow}=t_0$, and $\varphi_{0,(m,(n+1)),(m,n),\uparrow\uparrow}=\varphi_{0,(m,(n+1)),(m,n),\uparrow\downarrow}=\varphi_{0,(m,(n+1)),(m,n),\downarrow\uparrow}=\varphi_{0,(m,(n+1)),(m,n),\downarrow\downarrow}=2\pi\beta$. And set all other coefficients not mentioned above to be zero, we can realize the tight binding model in Eq. (\ref{hami}).
%Formula version
In $x$-direction, choose
\begin{equation} \label{para1}
\begin{aligned}
&t_{0,((m+1),n),(m,n),\uparrow\uparrow}=t_{0,((m+1),n),(m,n),\downarrow\downarrow}=t_0,\\ &\varphi_{((m+1),n),(m,n),\uparrow\uparrow}=2n\pi\alpha,\\
&\varphi_{((m+1),n),(m,n),\downarrow\downarrow}=-2n\pi\alpha.\\
\end{aligned}
\end{equation}
And in $y$-direction, set
\begin{equation} \label{para2}
\begin{aligned}
&t_{0,(m,(n+1)),(m,n),\uparrow\uparrow}=t_{0,(m,(n+1)),(m,n),\uparrow\downarrow}\\
=&t_{0,(m,(n+1)),(m,n),\downarrow\uparrow}=t_{0,(m,(n+1)),(m,n),\downarrow\downarrow}=t_0,\\ &\varphi_{0,(m,(n+1)),(m,n),\uparrow\uparrow}=\varphi_{0,(m,(n+1)),(m,n),\uparrow\downarrow}\\
=&\varphi_{0,(m,(n+1)),(m,n),\downarrow\uparrow}=\varphi_{0,(m,(n+1)),(m,n),\downarrow\downarrow}=2\pi\beta.
\end{aligned}
\end{equation}
When all other coefficients not mentioned above are set to be zero, we can realize the tight binding model in Eq.~(\ref{hami}). Equations~(\ref{para1}) and (\ref{para2}) show that $t_0$ can be adjusted by the corresponding hopping strength and $(\alpha,\beta)$ can be varied through changing hopping phases.

In experiments, adding detuning to the transition frequencies between the nearest-neighbor lattice in $y$-direction can also simulate staggered on-site potential. We take the column 1-3-1 to illustrate how the detuning is added as in Figs.~\ref{fig1}(c) and \ref{fig1}(d). And the hopping strength can be controlled by varying the amplitudes of each unit cell. The hopping phase can be adjusted by the SQUIDs between each lattices, which can drive phase transition between topological and trivial phases. That Peierls phase is hard to implement in cold atom systems because of the limitation that $\beta$ can not be small values \cite{Hamiltonian1}.

The validity of individual frequency addressing of the inter-cell transitions can be verified
%in our previous work, see Ref.\cite{TS1D_2}.  In addition, the
by numerical simulation. We find that in the present of the unmatched driving, all the initial non-target states remain almost unchanged, thus justify our individual frequency addressing method. So far, we have shown how to realize 2D tight-binding model by the combined TLRs and transmons system, next we will use this system simulate the 2D lattice in Fig.~\ref{fig1}(a).

\begin{figure}[tbp]
\center
\includegraphics[width=\linewidth]{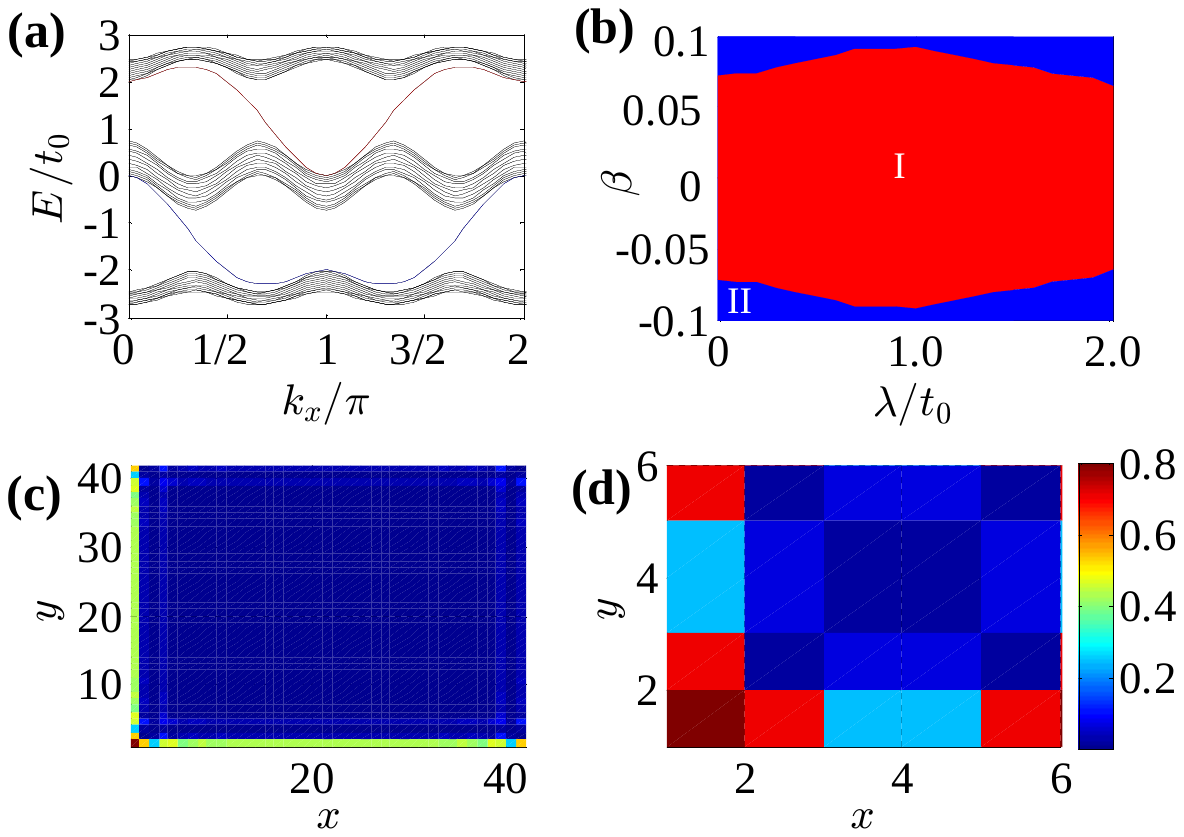}
\caption
{ (a) Energy band for $\lambda=0$ without spin-mixing $\beta=0$. (b) Phase diagram in $\beta-\lambda$ plane, based on the band structure in (a) with Fermi energy between $t_0$ to $2t_0$.
The plane is divided into two kinds: (I). topological insulator with edge states which is shown the so called QSHE; (II). metal state.
(c), (d) Distributions of the edge states' wave functions for (c) 42$\times$42 and (d) 6$\times$6 lattices.} \label{fig2}%for adding figures\
\end{figure}

\section{Verification of the Simulation}

Taking periodic boundary condition in $x$-direction, we numerically calculate the topological invariant \cite{TV1,TV2,review_band_calculation} and fix phase diagrams with corresponding Fermi energy, using the methods for analysing the $\mathbb{Z}_2$ topological insulators \cite{Kane_Z2,z2_analysis}.  And for real lattice case, we take open boundary condition. In the numerical work we set the hopping coefficient $t_0$ as energy unit, lattice spacings $a,b=1$ and $\alpha=1/3$.

\begin{figure}[ptb]
\centering
\includegraphics[width=\linewidth]{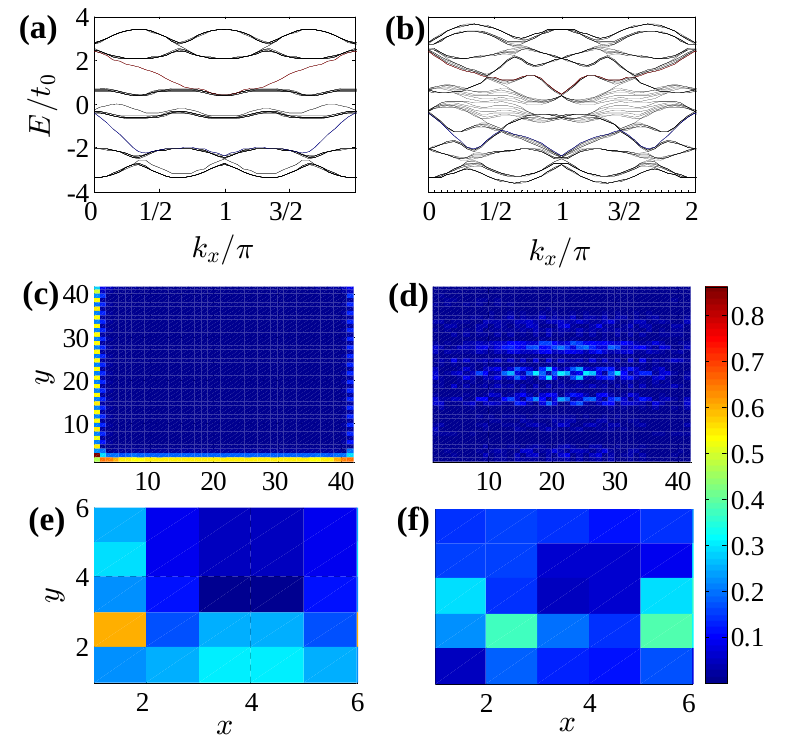}
\caption{(a), (b) Band structures for $\lambda=t_0$ with different spin-mixing terms for (a) $\beta=0$ and (b) $\beta=0.1$. (c), (e) The wave functions of the topological state with the same parameters as (a). (d), (f) The wave functions of the metal state with the same parameters as (b). (c,d) are the results for 42$\times$42 lattice, while (e,f) are the 6$\times$6 case.}\label{fig3}
\end{figure}

%\subsection{No spin-mixing case: $\beta=0$}
When $\beta=0$ there is no spin-mixing, and if $\lambda=0$ the Hamiltonian in Eq.~(\ref{hami}) are just two Hofstadter models with different magnetic fluxes $\pm2\alpha\pi$ for spin-up and spin-down branches. With the parameters $\lambda=0$ and $\beta=0$ the band structure of the system is plotted in Fig.~\ref{fig2}(a). With the numerical results of topological invariants, we can get Fig.~\ref{fig2}(b), which is the phase diagram with Fermi energy between $t_0$ to $2t_0$. The phase is divided into two kinds: topological and metal states. Based on this phase diagram we will show the effective quantum simulation of the QSHE in our system. We did numerical calculation and the wave functions of spin up topological state for different lattice sizes are shown in Figs.~\ref{fig2}(c) and \ref{fig2}(d). Considering the symmetry of the system, in this paper we only show the wave function of spin up state.
Along ~$\beta=0$ axis driving staggered potential $\lambda$ from $0$ to $t_0$, there is no phase transition but the QSHE will be affected and the wave distribution will change, such variation can be found in Figs.~\ref{fig2}(c,d) and \ref{fig3}(c,e).  And the wave function distributions on the surface in Figs.~\ref{fig2}(c) and \ref{fig3}(c) show the edge states along the boundary, which is just the perfect simulation of the QSHE. There is size effect in numerical calculation which we will discuss latter and we found a 42$\times$42 lattice is sufficient to cover all details of the physical system.

%\subsection{Spin-mixing case: $\beta\ne0$}
We next show how to trigger phase transition between topological insulator and metal state by changing the coupling parameter $\beta$ between spin-up and spin-down terms. In our scheme, adjusting the SQUIDs between lattices along $y$-direction can vary $\beta$ in Eq.~(\ref{hami}). Different from cold atom case \cite{Hamiltonian1}, it is not necessary to set $\beta$ to be a large value which means spin-up and spin-down states have to be mixed deeply. Look at Figs.~\ref{fig2}(c) and \ref{fig2}(d), when ~$\lambda=0$,  no spin-mixing case $\beta=0$, the system is topological insulator and if the spin-mixing efficient became larger the system will transform to metal state at last. We choose $\lambda=t_0$, $\beta=0$ and $0.1$ to show that phase transition.  Figures~\ref{fig3}(a) and \ref{fig3}(b) are the band structures for different spin-mixing terms $\beta=0$ and $\beta=0.1$ with $\lambda=t_0$. And in the same column are the wave functions of the corresponding spin up state for systems of 42$\times$42 and 6$\times$6 lattices. In Figs.~\ref{fig3}(c) and \ref{fig3}(e),  topological invariant $\nu=1$ and edge states appear along the boundary of the system. While Figs.~\ref{fig3}(d) and \ref{fig3}(f) show the wave functions for the metal states.

\section{Discussion}
Comparing (c), (e) and (d), (f) in Fig.~\ref{fig3}, predictably we see the size effects in these new results in Figs.~\ref{fig3}(e) and \ref{fig3}(f). However, when $\lambda=0$ and $\beta=0$, see Figs.~\ref{fig2}(c) and \ref{fig2}(d), the QSHE is not changed so much as $\lambda=t_0$ and $\beta=0$ case.  After numerical calculation, we find  $m,n\ge6$ is good enough to realize the QSHE, and of course  the more the better.  And mark $\alpha=1/q$, where q is an integer, set $n$ be an integral multiple of $q$ can effectively reduce the size effects. Considering the real status of experiments $n=6$ can be a good choice for the present $\alpha=1/3$.

\begin{figure}[tpbh]
\center
\includegraphics[width=0.5\linewidth]{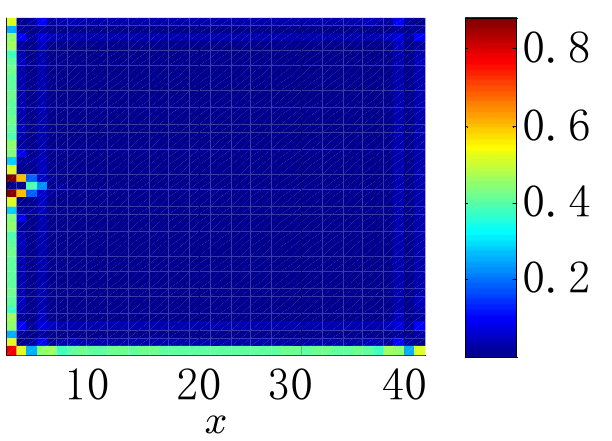}
\caption
{
(a). Topological edges state on the 42$\times$42 lattices with $\alpha=1/3, \beta=0, \lambda=0$. (b). Disconnect the lattice at (1,21) with its neighbours, and the same parameters are used in calculation. Numerical result shows that the edge modes steer by the impure lattice.
 \label{impurity}%for adding figures\
 }
\end{figure}

With the parameters of the edge states shown in Figs.~\ref{fig2}(c) and \ref{fig2}(d), the frequency and hopping coefficient of all the lattices include the lattices in the bulk must be set following the requirement in Eqs. (\ref{para1}) and (\ref{para2}), then the 2D system is in topological phase. Actually, the system has many eigen energies and eigen states in topological phase, we just choose one of them from the energy band for detection and label it as $E_{topo}$. Then prepare an original spin up state $|\uparrow\rangle_{\textbf{r}}$ with energy around $E_{topo}$ of the edge states. In experiment, after setting the parameters as theoretical proposal, the initialization of the system can be achieved by setting an edge transmon in its excited state $E_{\textbf{r},\uparrow}= \omega_{\textbf{r}} + g_{\textbf{r}}=E_{topo}$. Then investigate the excitation signal on the 2D superconducting quantum circuits with the predicted energy $E_{topo}$. The bulk states and the edge states have different eigen energies, so they can’t get populated at the same time, so the signal will only be detected in the edge lattices, that's simulated quantum spin Hall effect. And we can show that topological edges states are backscattering-immune with impurity\cite{prl_im,nature_im}. As contrast we take the topological edge state in Fig. 2(c) of the paper and add an impurity by disconnecting it with neighbor lattices by setting the hopping coefficients vanished in numerical calculation\cite{Xue_npj}. Comparing Fig. 2(c) and Fig. \ref{impurity}, we can see the topological edge states steer by the impure lattice without backscattering. That robustness of transportation is originated from the time reversal symmetry of the system, which is an important property of topological insulator. In experiment, the impurity can be set by turning off the connection between the impure lattice and its neighbors.
 Considering the progress of the experiments  about the superconducting circuits, there is limitation of the lattices' size in experiment for realizing the QSHE. As shown in Fig.~\ref{fig2}(d), 6$\times$6 lattice is enough to find the edge state, and can be achieved experimentally soon \cite{exp1,exp2,exp3,exp4}.

\begin{figure}[tbp]
\center
\includegraphics[width=0.8\linewidth]{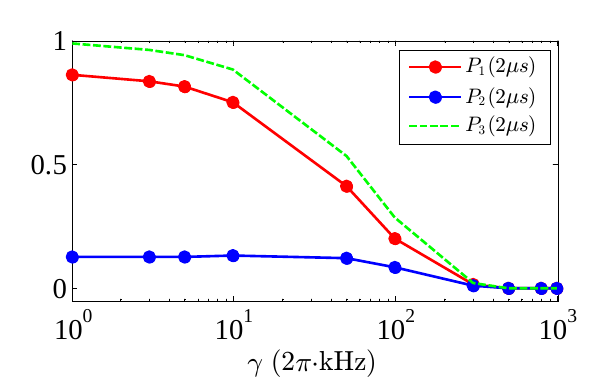}
\caption
{The edge-site population $P_1(t)$, the inter-site population $P_2(t)$ and the total population of the system $P_3=P_1+P_2$ versus the decay rate $\gamma$ for 6$\times$6 lattice in Fig.~\ref{fig2}(d) for the initial state $|\uparrow\rangle_{{\textbf{r}}}$ at ${\textbf{r}}=(1,1)$. All the results take evolution time 2$\mu$s.} \label{fig4}%for adding figures\
\end{figure}

As an addition, we investigate the quantum decoherence effects in our proposal for detecting the edge states. We use the Lindblad master equation and take three main decoherence factors in numerical calculation: the losses of the photon, the decay and dephasing of the transmon into account. The Lindblad master equation can be written as
\begin{equation}
    \dot {\rho }=-{\text{i}}[\mathcal{H}_{\text{JC}},\rho ]+\sum_{\textbf{r}}\sum _{i=1}^{3}\gamma\left(\Gamma_{{\textbf{r}},i}\,\rho \Gamma_{{\textbf{r}},i}^{\dagger }-{\frac {1}{2}}\left\{\Gamma_{{\textbf{r}},i}^{\dagger }\Gamma_{{\textbf{r}},i},\rho \right\}\right),
\end{equation}
where $\rho$ is the density operator of the system, $\gamma$ is the decay rate or noise strength which are set to be the same here, $\Gamma_{{\textbf{r}},1}=a_{{\textbf{r}}},\;\Gamma_{{\textbf{r}},2}=\sigma^-_{{\textbf{r}}}$ and $\Gamma_{{\textbf{r}},3}=\sigma^z_{{\textbf{r}}}$ are the photon-loss, transmon-loss  and the transmon-dephasing operators in the lattice at $\textbf{r}$, respectively. In Fig.~\ref{fig4}, we plot the edge-site and inner-site populations:
 \begin{equation}
 \begin{aligned}
P_1(t)&=\text{tr}[\rho(t)\sum_{{\textbf{r}_{\text{edge}}}}(a_{\textbf{r}}^\dag a_{\textbf{r}}+\sigma_{\textbf{r}}^+\sigma_{\textbf{r}}^-)],\\
  P_2(t)&=\text{tr}[\rho(t)\sum_{{\textbf{r}_{\text{inner}}}}(a_{\textbf{r}}^\dag a_{\textbf{r}}+\sigma_{\textbf{r}}^+\sigma_{\textbf{r}}^-)],
\end{aligned}
\end{equation}
   after 2$\mu$s for different decay rates. It shows that both the edge state population $P_1$ and the inner state population $P_2$ decrease smoothly when the decay rate increases. Fortunately, the detection in our scheme can tolerate the decay rate up to the order of $2\pi\times 10$ kHz, while the typical decay rate is $2\pi\times 5$ kHz. We use the initial state $|\uparrow\rangle_{{\textbf{r}}={(1,1)}}$ in the numerical calculation in the consideration that it is easier to preparing such excited state in one site than the eigenstate of the Hamiltonian in Eq.~(\ref{hami}) which concerns all 36 sites on the 2D plane. That initial state caused a little leakage from edge state to inner state, however that not effects much in the detection.

\section{Conclusion}

In summary, we have proposed a circuit quantum electrodynamics system with TLRs and transmons connected by SQUIDs,
the hopping coefficients between these lattices can be adjusted separately. With that a $\mathbb{Z}_2$ topological insulator
in 2D lattices is realized and the phase transitions between topological and trivial states are simulated. Our proposal is stable and well controllable, especially using SQUIDs to realize
the hopping is smooth and steady. The state of each lattice can be prepared and detected separately and precisely. And superconducting circuits can be used to simulate gauge fields effect that had been done with cold atoms and still keep the present advantages. These new characteristics surpass the previous methods and will shed light not only on the realization of the topological systems but also topological quantum computation.

\section*{Acknowledgments}
This work was supported by the Key-Area Research and Development Program of Guangdong province (Grant No.~2018B030326001), the National Natural Science Foundation of China (Grant No.~11874156, No.~11904111, and No.~11674200), the National Key R\&D Program of China (Grant No.~2016YFA0301803), the Project funded by China Postdoctoral Science Foundation (Grant No.~2019M652684), and the Science and Technology Program of Guangzhou (Grant No.~2019050001).

The authors declare no conflict of interest.

\end{document}